\documentclass[preprint,showpacs,aps,amssymb,floatfix,prd,amsmath,preprintnumbers,showkeys]{revtex4} 
\usepackage{epstopdf}
\usepackage{capt-of}
\usepackage{graphicx}  
\usepackage{dcolumn}   
\usepackage{bm}
\begin{document}
\input epsf.tex

\title{Effective gravitational mass of the Ay\'{o}n-Beato and Garc\'{\i}a metric\footnote{\scriptsize Most  of this work was done during an International  Workshop on  Introduction to Research in Einstein's General Relativity at NIT, Patna (India). Authors' emails as well as their  permanent addresses are mentioned below:}}

\author{
        A. K. Sinha\footnote{\scriptsize{Department of Physics, College of Commerce, Patna 800020, Bihar, India, Email: ashutosh25june@gmail.com}},
	G. K. Pandey\footnote{\scriptsize{Patna Science College, Patna University, Patna 80005, Bihar, India, Email: gaurav.golu1@gmail.com}},
	A. K. Bhaskar\footnote{\scriptsize{Department of Physics, College of Commerce, Patna 800020, Bihar, India, Email: drakbhaskar@gmail.com}},	
	B. C. Rai\footnote{\scriptsize{Department of Physics, College of Commerce, Patna 800020, Bihar, India, Email: bcraiphy@gmail.com}},	
	A. K. Jha\footnote{\scriptsize{Department of Physics, College of Commerce, Patna 800020, Bihar, India, Email: drarunkjhacoc@gmail.com}},
	S. Kumar\footnote{\scriptsize{Department of Physics, College of Commerce, Patna 800020, Bihar, India, Email: skphysics@yahoo.co.in}}}
\affiliation{National Institute of Technology, Patna 800005, Bihar, India. \\  }

\author {S.~S.~Xulu\footnote{\scriptsize{ Email: xuluss@unizulu.ac.za}}}
    \affiliation{Department of Computer Science, University of Zululand,3886 Kwa-Dlangezwa, South Africa. \vspace*{0.5in}}


\begin{abstract}
In this paper, we  calculate  the effective gravitational mass of Ay\'{o}n-Beato and Garc\'{\i}a  regular (non-singular) static spherically symmetric asymptotically Minkowskian  metric that is a solution  to Einstein's equations coupled with a nonlinear electromagnetic field. The effective gravitational mass is negative, zero, or positive that depends on the  ratio of magnitude of electric charge to the ADM mass and the ratio of the radial distance to the ADM  mass. As expected, at large value of radial distance, our result gives effective gravitational mass of  the Reissner-Nordstr\"om  metric.

\end{abstract}




\maketitle


\section{Introduction}
In the year 1984, Cohen and de Felice\cite{CF84} defined effective gravitational mass of a metric at any point in a given 
space-time  at a radial distance $r$ as the energy contained inside  the region of $r=constant$ surface,  and  used this to explain repulsive nature of timelike singularities. They used Komar energy formula and obtained effective  gravitational  mass of the Kerr-Newman metric (characterized by mass $M$, electric charge $q$, and rotation parameter $a$):
\begin{equation}
\mu(r)= M-\frac{q^2}{2 r}
\left[
 1 + \frac{\left(a^2+r^2\right)}
              { a r} \arctan\left(\frac{a}{r}\right)
\right] \text{.}
\label{CohenKN}
 \end{equation}
For large values of rotation and/or charge parameters (that is,  when these parameters dominate over the mass parameter), the effective gravitational mass becomes negative.  For the  Reissner-Nordstr\"om  metric ($a=0$) in the above result, the effective gravitation mass  is
\begin{equation}
\mu(r)= M-\frac{q^2}{r} \text{.}
\label{CohenRN}
 \end{equation}
Obviously, $\mu(r) < 0$ for $r < \frac{q^2}{M}$ and this explains repulsive gravitational  effect on an electrically neutral test particle in the gravitational field described by the Reissner-Nordstr\"om  metric. This fascinating interpretation of energy content in a space-time attracted researchers; however, they soon realized that there is no unique adequate formalism for energy calculation and this had been a difficult problem since the appearance of general theory of relativity.  For example, Komar definition   cannot be applied to non-static  space-times.

Though Einstein's theory of gravity is a well governing theory of space, time, and gravitation and it is  very well  experimentally verified, and is, so far, proven to be the most successful theory to describe gravity at small as well as large scales, it does possess some issues. The energy-momentum localization problem is one such important issue in the context of General Relativity. Energy, momentum, and angular momentum are important conserved quantities in Minkowski space-time  (when gravitational field is  negligibly small.) These have significant role as they provide the first integrals of motion to solve otherwise  unmanageable physical problems. However, till today, we do not have a general definition to describe energy-momentum  distribution in curved space-times (i.e., in presence of gravitational field) and several difficulties arouse in this direction. 

All the attempts to finding this  resulted in a large number of different energy-momentum complexes. To solve this problem, Einstein  formulated the energy-momentum local conservation law (see in \cite{Moller}). Then, many physicists including, Landau-Lifshitz\cite{LL},  Weinberg\cite{Weinberg}, Papapetrou\cite{Papapetrou}, and Bergman and Thomson\cite{BT} proposed various definitions of energy-momentum distribution. But, these definitions of energy-momentum complexes were coordinate dependent, that is, they give meaningful results only when the calculations were done in quasi-Cartesian coordinates. It was suspected that a plethora of different energy-momentum complexes would give acceptable total energy and momentum for isolated systems (i.e., asymptotically flat space-times); however, they would  produce different and hence meaningless energy-momentum distributions  even in asymptotically flat (Minkowskian) spacetime and  would not give any  meaningful result at all for asymptotically non-Minkowskian spacetimes. Virbhadra's and his collaborators\cite{KSV1,KSV2} seminal work  shook this prevailing prejudice and they explicitly showed that different energy-momentum complexes give the same results for many spacetimes (asymptotically Minkowskian as well as non-Minkowskian). Later, Virbhadra and Rosen\cite{RV}  (the most famous collaborator of Albert Einstein)  studied Einstein-Rosen gravitational waves and for this metric as well different complexes gave same and  reasonable results. Further,  attracted by these encouraging results, many  researchers\cite{M1,M2,M3} studied several other metrics and obtained very useful results.

In General Theory of Relativity, majority of  well-known exact black holes solutions to the Einstein's field equations came up with the  existence of space-time (curvature) singularities where general relativity theory breaks down. Therefore, it is very desirable to have regular (nonsingular) solutions of Einstein's equations.
To avoid this black hole curvature singularity aspect, researchers came up with a large number of regular black hole models which were referred to as ``Bardeen black holes''\cite{Bardeen} because Bardeen was the first to obtain a regular black hole solution. But, the problem with all of them was that none of them were an exact solution to the Einstein's Field Equations. Moreover, there are no known physical sources pertaining to  any of those regular black hole solutions.

Later in 1998, Ay\'{o}n-Beato and Garc\'{\i}a found out the first  singularity free exact black hole solution\cite{AG}  to the Einstein's field equations whose source is a non-linear electrodynamic field  coupled to gravity and the metric does not possess any curvature singularity. In this paper, we calculate effective gravitational mass of this metric and analyze the results.  We use  M{\o}ller's \cite{Moller}  energy-momentum expression which permits  to calculate the energy-momentum distribution in an arbitrary  coordinate system.  As usual in  general relativity papers, we  too use the geometrized units ($G=1,c=1$).  We use Mathematica \cite{Math} software  for plots.


\section{The Ay\'{o}n-Beato and Garc\'{\i}a solution}
The Ay\'{o}n-Beato and Garc\'{\i}a   (hereafter referred to as AG) static spherically symmetric asymptotically Minkowskian solution to Einstein's equations with a  source of nonlinear electrodynamic field  is given by the  line-element
\begin{equation}
ds^2=\beta dt^2- \alpha dr^2-r^2 (d\theta^{2}+\sin^{2}\theta d\phi^2)
\label{AGSch}
\end{equation}
and the electric field
\begin{equation}
E = q r^4 
      \left(
      \frac{r^2-5q^2}{\left(r^2+q^2\right)^4}  
       + \frac{15}{2} \frac{M}{\left(r^2+q^2\right)^{7/2} }
      \right) \text{,}
\end{equation}
where
\begin {equation}
\beta = \frac{1}{\alpha} = 1-\frac{2M r^2}{(r^2+q^2)^\frac{3}{2}} + \frac{q^2 r^2}{(r^2+q^2)^2} \text{.}
\end{equation}
Ay\'{o}n-Beato and Garc\'{\i}a \cite{AG} showed that their solution corresponds to charged non-singular  black holes for  $|q| \leq$  approximately $0.6 M$ and otherwise the solutions are though regular, there are no horizons.

The asymptotic behavior  of the AG solution is given by 
\begin{eqnarray}
\alpha & = &  1+\frac{2 M}{r}+\frac{4 M^2-q^2}{r^2}+O\left(\frac{1}{r^3}\right) \text{,} \nonumber \\ 
\beta  & = &  1-\frac{2 M}{r}+\frac{q^2}{r^2}+O\left(\frac{1}{r^3}\right) \text{, and} \nonumber \\
E     &  = & \frac{q}{r^2}+O\left(\frac{1}{r^3}\right) \text{.} 
\end{eqnarray}
Thus the A-G solution behaves asymptotically  as the Reissner-Nordstr\"om solution to Einstein-Maxwell equations. 
The symbol $M$ represents the  ADM mass and  $q$  stands for the electric charge. 
The behavior of the  above functions  near $r=0$:
\begin{eqnarray}
\alpha & = &  1+r^2 \left(\frac{2 M |q|}{q^4}-\frac{1}{q^2}\right)+O\left(r^3\right) \text{,} \nonumber \\ 
\beta  & = &  1-r^2 \left(\frac{2 M |q|}{q^4}-\frac{1}{q^2}\right)+O\left(r^3\right) \text{,} \nonumber \\
E     &  = &  \left(\frac{15 |q|}{2 q^6}-\frac{5}{q^5}\right) r^4+O\left(r^6\right)\text{.} 
\end{eqnarray}
Thus, the metric behaves as de Sitter near $r=0$.
As we will  need the metric in quasi-Cartesian coordinate system as well, applying the coordinate transformation
\begin{eqnarray}
x &=& r \sin \theta \cos \phi \text{,} \nonumber \\ 
y &=& r \sin \theta \sin \phi  \text{,} \nonumber \\
z &=& r \cos \theta \text{,} 
\end{eqnarray}
the line element gets  the form,
\begin{equation}
ds^2=\beta dt^2-(\gamma x^2+1)dx^2-( \gamma y^2+1)dy^2-( \gamma z^2+1) dz^2-2 \gamma(x y dx dy + y z dy dz + z x dz dx) \text{,}
\label{AGCart}
\end{equation}
where
\begin{equation}
 \gamma =\frac{A-1}{x^2+y^2+z^2} \text{.}
\end{equation}
 

\section{M{\o}ller Energy-Momentum Prescription}

The M{\o}ller energy-momentum complex is given by \cite{Moller}

\begin{equation}
\Im_i^{kl} = \frac{1}{8\pi} \chi_i^{kl}{, l} \; ,
\end{equation}

which satisfies the local conservation laws as

\begin{equation}
\frac{\partial{\Im_i^{k l}}}{\partial{x^k}} = 0 \; ,
\end{equation}

where the antisymmetric M{\o}ller super-potential $\chi_i^{k l}$ is

\begin{equation}
\chi_i^{k l} = -\chi_i^{l k } = \sqrt{-g} \ [g_{in,m}-g_{im,n}]\ g^{k m}g^{n l}  \;.
\end{equation}

The energy and  momentum components are  now given by

\begin{equation}
P_i=\int \int \int \Im_i^0 \ dx^1 dx^2 dx^3 \;.
\end{equation}

$P_i$ stands for momentum components $P_1$, $P_2$, $P_3$, and $P_0$ is the energy. Following Cohen and de Felice \cite{CF84}, we interpret the energy function as the effective gravitational mass $\mu(r)$ of the metric. We apply  Gauss' theorem in above equation and then   the total energy and momentum components  take  the new form:
\begin{equation}
P_i=\frac{1}{8\pi} \int \int \chi^{0\delta}_{i} \ n_{\delta} \ dS \;,
\label{Pi}
\end{equation}
where $n_{\delta}$ is the outward unit normal vector over an infinitesimal surface element $dS$.
\section{Calculations}

M{\o}ller claimed that his energy-momentum complex can be used in any coordinate system and would produce the same results. In order to  obtain the effective gravitational mass of the AG metric and to verify M{\o}ller's claim, we  first perform calculations in Schwarzschild coordinates $\{t,r,\theta, \phi\}$ and  then in quasi-Cartesian coordinates $\{t,x,y,z\}$.

\subsection{Calculations in Schwarzschild  coordinates}
In Schwarzschild  coordinates, the determinant of the covariant metric tensor $g_{ik}$ for the AG metric given  by Eq. $(\ref{AGSch})$ is  
\begin{equation}
g = -r^4 \sin^{2} \theta \text{.}
\end{equation}
As the metric is represented by a diagonal matrix, $g^{ik} = \frac{1}{g_{ik}} \forall$ values of indices $i$ and $k$.  The only  non-vanishing component of $\chi_i^{k l}$  which is needed for the energy calculation is
\begin{equation}
\chi_0^{0 1} = - \chi_0^{1 0} =   2 r^3\sin\theta\biggl[M \frac{r^2- 2q^2}{(r^2+q^2)^\frac{5}{2}}- q^2 \frac{r^2-q^2}{(r^2+q^2)^3}\biggr]
\end{equation}
while all the other components of $\chi_i^{k l}$ vanishes.
Using the above expression in equation $(\ref{Pi})$, the energy distribution is given by

\begin{equation}
{\cal E}_{S}= r^3\biggl[M\frac{r^2-2q^2}{(r^2+q^2)^\frac{5}{2}}-q^2\frac{r^2-q^2}{(r^2+q^2)^3}\biggr]
\label{ENSch}
\end{equation}

The subscript $S$ emphasizes that calculations have been performed using  Schwarzschild coordinates $\{t, r, \theta, \phi\}$. Thus, we find the same result for M{\o}ller energy in Schwarzschild coordinates as obtained by  Yang et al. in \cite{Yangetl1}. (The same research team also obtained  M{\o}ller energy in Schwarzschild coordinates for a more general metric \cite{Yangetl2}.) Now we calculate 3 momentum components and all are found to be zero, i.e.,
\begin{equation}
P_r = P_{\theta} = P_{\phi} = 0 \text{,}
\label{MomSch}
\end{equation}
and this is desired result as the AG metric is static.
\subsection{Calculations in quasi-Cartesian Coordinates}
The desired covariant components of fundamental metric tensor  $g_{ik}$ is obtained from the equation $(\ref{AGCart})$: 

\begin{equation}
g_{ik} = \left(
\begin{array}{cccc}
 \beta  & 0 & 0 & 0 \\
 0 & -\gamma  x^2-1 & -x y \gamma  & -x z \gamma  \\
 0 & -x y \gamma  & -\gamma  y^2-1 & -y z \gamma  \\
 0 & -x z \gamma  & -y z \gamma  & -\gamma  z^2-1 \\
\end{array}
\right)
\end{equation}
The determinant of this metric tensor is,
\begin{equation}
g = |g_{ik}| = -1 \text{.}
\end{equation}

The  components of $g^{ik}$ are given by 

\begin{equation}
g^{ik} = 
 \left(
\begin{array}{cccc}
 \frac{1}{\beta } & 0 & 0 & 0 \\
 0 & -\beta  \left(\text{$\gamma $y}^2+\text{$\gamma $z}^2+1\right) & x y \beta \gamma  & x z \beta \gamma  \\
 0 & x y \beta  \gamma  & -\beta  \left(\text{$\gamma $x}^2+\text{$\gamma $z}^2+1\right) & y z \beta \gamma  \\
 0 & x z \beta  \gamma  & y z \beta  \gamma  & -\beta  \left(\text{$\gamma $x}^2+\text{$\gamma $y}^2+1\right) \\
\end{array}
\right) \text{.}
\end{equation}
The components of the super-potential $ \chi_i^{k l}$ in quasi-Cartesian coordinates are obtained here: 
\begin{eqnarray}
\chi_0^{01}  &=&  \frac{\partial{\beta}}{\partial{x}}\text{,}\nonumber \\ 
\chi_0^{02}  &=&  \frac{\partial{\beta}}{\partial{y}} \text{,}\nonumber\\ 
\chi_0^{03}  &=&  \frac{\partial{\beta}}{\partial{z}} \text{,} \nonumber\\ 
\chi_1^{01}  &=&  \chi_1^{02} = \chi_1^{03} = 0  \text{,} \nonumber\\
\chi_2^{01}  &=& \chi_2^{02} =\chi_2^{03} = 0 \text{,} \nonumber\\
\chi_3^{01}  &=& \chi_3^{02} = \chi_3^{03} = 0 \text{.}
\end{eqnarray}
Using these $\chi_i^{jk}$ into the equation (\ref{Pi}), we obtain the energy-momentum components in quasi-Cartesian coordinates,
\begin{equation}
{\cal E}_{QC}= r^3\biggl[M\frac{r^2-2q^2}{(r^2+q^2)^\frac{5}{2}}-q^2\frac{r^2-q^2}{(r^2+q^2)^3}\biggr]
\label{ENQC}
\end{equation}

The subscript $QC$ emphasizes that calculations have been performed using  quasi-Cartesian coordinates $\{t, x, y, z\}$. All the three momentum components are found to be zero, i.e.,
\begin{equation}
P_x = P_{y} = P_{z} = 0 \text{,}
\label{MomQC}
\end{equation}

M{\o}ller energy in Schwarzschild coordinates  obtained by Yang et al \cite{Yangetl1} and our results, given by equations  $(\ref{ENSch})$,  $(\ref{MomSch})$, $(\ref{ENQC})$, and  $(\ref{MomQC})$ for energy and momentum distributions in AG space-time, support the claim of M{\o}ller  that his definition of energy and momentum is applicable  to  both Schwarzschild coordinates  $\{t, r, \theta, \phi\}$ and quasi-Cartesian coordinates  $\{t, x, y,z\}$.

\subsection{Effective gravitational  mass of the AG metric}

In the previous sub-section, we showed that M{\o}ller's definition of energy-momentum produces coordinate-independent results. Now, following Cohen and de Felice\cite{CF84}, we interpret the energy distribution as the effective gravitational mass $\mu(r)$,  and then thoroughly analyze the results and explain its importance. 

The effective gravitational mass of the AG metric is

\begin{equation}
{\mu(r)} = r^3\biggl[M\frac{r^2-2q^2}{(r^2+q^2)^\frac{5}{2}}-q^2\frac{r^2-q^2}{(r^2+q^2)^3}\biggr]
\label{EGM}
\end{equation}

The asymptotic behavior  of the effective gravitational mass is
\begin{equation}
{\mu(r)} = M-\frac{q^2}{r}-\frac{9 M q^2}{2 r^2}+\frac{4 q^4}{r^3}+\frac{75 M q^4}{8 r^4}-\frac{9
   q^6}{r^5}+O\left(\frac{1}{r^6}\right) \text{.}
\label{AsyEGM}
\end{equation}
Thus the asymptotic value of the effective gravitational mass is the ADM mass $M$. It is also clear that at large $r$, the effective gravitational mass of the AG metric becomes the effective gravitational mass of the Reissner-Nordstr\"om  metric as obtained by Cohen and de Felice\cite{CF84}  using Komar energy and by Virbhadra\cite{KSV1}  using M{\o}ller's complex. $q=0$ in the above equation gives the effective gravitational mass of the Schwarzschild metric.

\begin{figure*}[tbh]
\includegraphics[width=1.0\linewidth]{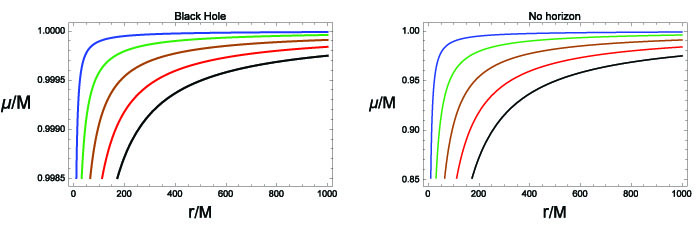}
\caption[ ]{(color online).
{In the figure on left side, the ratio of the  effective gravitational mass $\mu$ to the ADM mass $M$  is plotted against the ratio of the radial distance $r$ to the ADM mass $M$ for $Q/M = 0.1 {\text {(blue)}}, 0.2 {\text {(green)}}, 0.3{\text {(orange)}}, 0.4{\text {(red)}}, {\text {and}}\  0.5 {\text {(black)}}$. Further, 
the figure on right side, the ratio of the  effective gravitational mass $\mu$ to the ADM mass $M$  is plotted against the ratio of the radial distance $r$ to the ADM mass $M$  for $Q/M = 1 {\text {(blue)}}, 2 {\text {(green)}}, 3{\text {(orange)}}, 4{\text {(red)}}, {\text {and}} 5 {\text {(black)}}$. Plots on left and right sides are, respectively, for regular charged massive black holes and for regular charges massive objects with no horizons.
}}
\label{fig1}
\end{figure*}


\begin{figure*}[tbh]
\includegraphics[width=1.0\linewidth]{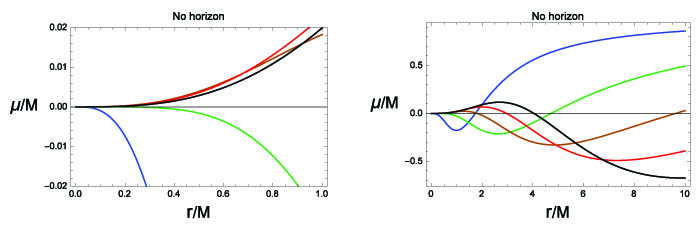}
 \caption[ ]{(color online).
{The ratio of the  effective gravitational mass $\mu$ to the ADM mass $M$  is plotted against the ratio of the radial distance $r$ to the ADM mass $M$ for $|q|/M = 1 {\text {(blue)}}, 2 {\text {(green)}}, 3{\text {(orange)}}, 4{\text {(red)}}, {\text {and}}\  5 {\text {(black)}}$ in the vicinity of the center.

}}
\label{fig2}
\end{figure*}


\begin{figure*}[tbh]
\includegraphics[width=1.0\linewidth]{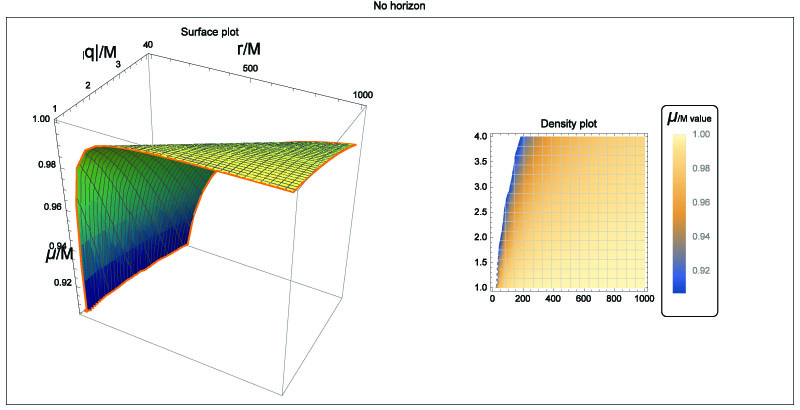}
 \caption[ ]{(color online).
{
The ratio of the  effective gravitational mass to the ADM mass, i.e., $\mu/M$  is plotted against the ratio of the radial distance  to the ADM mass, i.e., $r/M$ and   the ratio of the charge to the ADM mass $|q|/M $. The figure on left  side is a surface plot whereas the one on right side is the corresponding density plot. This shows  behavior of $\mu/M$ at large distances.

}}
\label{fig3}
\end{figure*}

\begin{figure*}[tbh]
\includegraphics[width=1.0\linewidth]{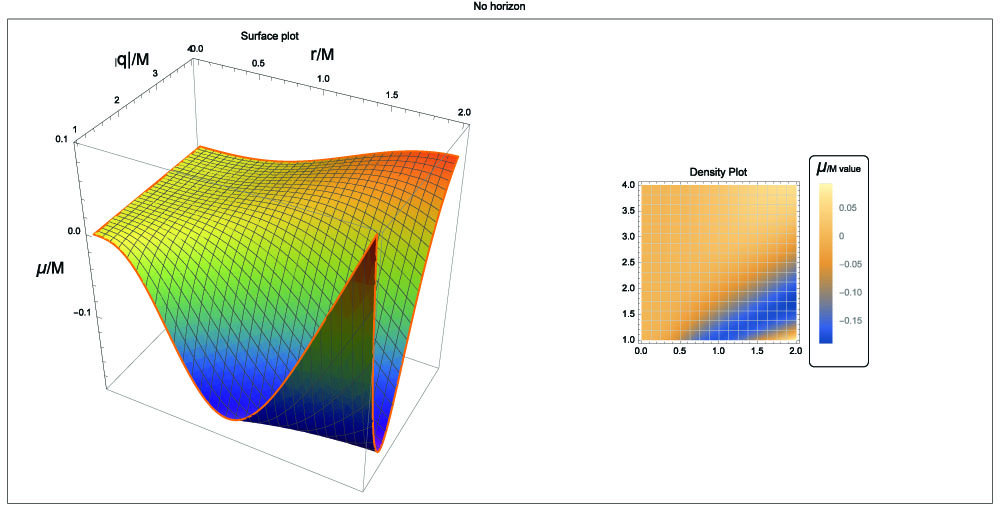}
 \caption[ ]{(color online).
{
The ratio of the  effective gravitational mass to the ADM mass, i.e., $\mu/M$  is plotted against the ratio of the radial distance  to the ADM mass, i.e., $r/M$, and  the ratio of the charge to the ADM mass $|q|/M $. The figure on left  side is a surface plot whereas the one on right side is the corresponding density plot. This plot shows the behavior of  $\mu/M$ close to the center.

}}
\label{fig4}
\end{figure*}

\begin{figure*}[tbh]
\includegraphics[width=1.0\linewidth]{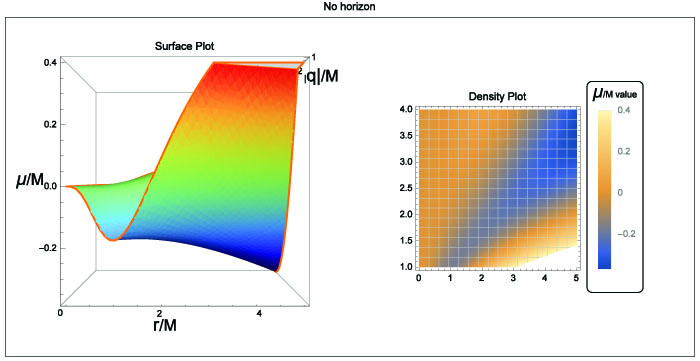}
 \caption[ ]{(color online).
{
The ratio of the  effective gravitational mass to the ADM mass, i.e., $\mu/M$  is plotted against the ratio of the radial distance  to the ADM mass, i.e., $r/M$ and the ratio of the charge to the ADM mass $|q|/M $. The figure on left  side is a surface plot whereas the one on right side is the corresponding density plot. This plot exhibits the behavior of  $\mu/M$  very close the center.
}}
\label{fig5}
\end{figure*}


The behavior of the effective gravitational mass near $r = 0$ is
\begin{equation}
{\mu(r)} = r^3 \left(\frac{1}{q^2}-\frac{2 M |q|}{q^4}\right)+r^5 \left(\frac{6 M
   |q|}{q^6}-\frac{4}{q^4}\right)+O\left(r^7\right) \text{.}
\label{NearZeroEGM}
\end{equation}
Thus, depending on the values of $\frac{|q|}{M}$  and $\frac{r}{M}$, the effective gravitational mass can be negative, zero, or positive. We now make numerous plots to see the behavior of the effective gravitational mass at large  as well as small  radial distances and also to see the facinating role  of $\frac{|q|}{M}$ on the effective gravitational mass.

In Figure 1, we plot the ratio of  the effective gravitational mass to the ADM mass  against the ratio of the radial distance to the ADM mass for different values of $|q|/M$. The asymptotic value for each case is $1$ showing that the total mass of the metric is the ADM mass $M$. For any fixed value of  $\frac{r}{M}$, the effective gravitational mass is smaller for higher value of $\frac{|q|}{M}$. Thus, the electric charge contributes negatively to the effect gravitational mass. As, for black holes,   $\frac{|q|}{M}$ is lower than for charged regular objects with no horizon, for any fixed value of  $\frac{r}{M}$ (but outside the even horizon), black holes have higher effective gravitational masses compared to  those without event horizon. In Figure 3, the same effects are shown more clearly  through surface and density plots.

In Figure 2, We plot the same quantities as we plotted in Figure 2; however, we plot  for no horizon cases only and near the center ($r=0$). Near the center, unlike the cases of large radial distances, higher  value of  $\mu/M$ for lower  value  $|q|/M$ does not always hold.  In fact, near $r=0$, $\mu/M$  could be  negative, zero, or positive and that is determined by both $|q|/M$ and $r/M$. The same quantities are plotted 
in figures 4 and 5 as  surface and density plots which show  with more clarity the variation of the effective mass to the ADM mass ratio as a function of the radial distance to the ADM  mass as well as the ratio of the absolute value of electric charge to the ADM mass.

\section{Summary}

The AG solution is a static spherically symmetric and  asymptotically Minkowskian regular solution of Einstein's equations 
coupled with a nonlinear electrodynamic field.  The solution behaves as  Reissner-Nordstr\"om solution  at a large distance  and 
de sitter  as r approaches zero.  For $|q| \leq$ approximately $0.6M$, this represents black holes and for 
larger values of $|q|/m$ there is no event horizon. 

We calculated the effective gravitational mass of the AG metric and extensively analyzed the results. At large distances, the role of the electric charge $|q|$ is to decrease the effective gravitational mass. However for small value of radial  distances, near $r = 0$, the effective gravitational mass dependence on  the electric charge  is not necessarily
similar and  the effective gravitational mass could be positive, zero, or negative depending on the radial distance to the ADM mass ratio  and relative magnitude  of absolute value of 
electric charge compared to the ADM mass. As  we already discussed, Yang et al \cite{Yangetl2} obtained M{\o}ller energy in Schwarzschild coordinates. Our investigation in this  work   evidences that M{\o}ller's prescription to find energy and momentum is unaffected by
the choice of different coordinate systems. 

The study of  effective gravitational mass that an electrically neutral test particle experiences predict 
many  physical effects even before calculations. A metric exhibiting negative  effective gravitational mass has repulsive    
effects not only to timelike  but  also to null geodesics, and hence also affect gravitational  lensing  phenomenon. 
Virbhadra's pioneer research \cite{KSVLens} in gravitational lensing as a tool to  propose astronomical test to the unproven cosmic censorship was influenced by his works and analysis of  the energy distribution (effective gravitational mass) in spacetimes. The knowledge of the effective  gravitational mass of AB metric would  give  more physical  insights about the spacetime which might have significant applications  to relativistic  astrophysics.

\section{Acknowledgments}
We are grateful to our mentor Dr. K. S. Virbhadra who taught us general relativity and introduced us to use software Mathematica, Reduce, and 
LaTex. He also introduced us to research in general relatvitiy. He cannot be thanked enough.
We also heartily thank the National Institute of Technology (NIT), Patna, India for organizing the International Winter Workshop on \textit{Introduction to Research in Einstein's General Relativity} during which this research work was done. We are also thankful to the people of  NIT Patna for their hospitality and the support.  SSX thanks the University of Zululand (South Africa) for all support.



\newpage

\begin{references}
\bibitem{CF84} J.~M.~Cohen and F.~ de Felice, J. Math. Phys., {\bf 25} 992 (1984); 
 A.~Komar, Phys.~Rev.~{\bf 113}, 934 (1959).
\bibitem{Moller} C.~M{\o}ller, Ann.~Phys. (NY) {\bf 4}, 347 (1958);  {\bf 12}, 118 (1961).

\bibitem{LL} L. D. Landau and E. M. Lifshitz, {\em The Classical Theory of Fields} 
         {Pergamon Press, 1987} p. 280.
\bibitem{Weinberg}
    S. Weinberg, {\em Gravitation and Cosmology:  Principles and  Applications of General Theory of Relativity} (John Wiley and 
     Sons, Inc., New York, 1972) p. 165.
\bibitem{Papapetrou} A. Papapetrou, Proc. R. Irish. Acad. {\bf A52}, 11   (1948).
\bibitem{BT} P. G. Bergmann and R. Thomson, Phys. Rev. 89, 400 (1953).
\bibitem{KSV1}
     K. S. Virbhadra,Phys. Rev. {\bf D41}, 1086 (1990); {\it ibid} {\bf D42}  1066 (1990); {\it ibid} 
     {\bf D42}, 2919 (1990);  A.~Chamorro and K.~S.~Virbhadra, Pramana-J.~Phys. {\bf 45}, 181 (1995); Int. J. Mod. Phys.        {\bf D5}, 251 (1996); K. S. Virbhadra and J. C. Parikh,   Phys. Lett. {\bf B317}, 312 (1993);
     {\it ibid} {\bf B331} 302 (1994) ; K.~S.~Virbhadra, Int .J. Mod. Phys. {\bf A12} 4831 (1997); {\it ibid} {\bf D6}           357 (1997); Pramana 44 317 (1995); Pramana {\bf 38} 31 (1992); Phys.Lett. {\bf A157}  195 (1991). 

\bibitem{KSV2} K. S. Virbhadra, Phys.\ Rev.\  {\bf D60}, 104041 (1999); J. M. Aguirregabiria, A. Chamorro  and  K. S.    
               Virbhadra, Gen. Relativ. \& Gravit. {\bf 28}, 1393 (1996).
\bibitem{RV}  N. Rosen and K. S. Virbhadra,   Gen. Relativ. Gravit. {\bf 25}, 429 (1993); 
                  K. S. Virbhadra,  Pramana 45 215 (1995). 


\bibitem{M1} S. S. Xulu, Int. J.Theor. Phys. {\bf 37}  1773 (1998); Int. J.Mod. Phys. {\bf D7}
             773 (1998); Int.\ J.\ Mod.\ Phys.\ A {\bf 15}, 2979 (2000); Int.\ J.\ Theor.\ Phys.\  
           {\bf  39}, 1153 (2000); Int.\ J.\ Mod.\ Phys.\ A {\bf 15}, 4849 (2000); Mod.\ Phys.\ Lett.\ A    {\bf 15}, 1511                 (2000); Astrophys.\ Space \ Sci. {\bf 283}, 23 (2003);S. S. Xulu,  Int.\ J.\ Theor.\ Phys.\  {\bf 46},           2915 (2007); Chin.\ J.\ Phys.\  {\bf 44}, 348 (2006); Found.\ Phys.\ Lett.\  {\bf 19}, 603 (2006); I.              Radinschi, Chin. J. Phys. \textbf{39},
           393 (2001); Chin. J. Phys. \textbf{39}, 231 (2001); Int. J. Mod. Phy. D \textbf{13}, 1019 (2004); 
   R.~M.~Gad and M.~F.~Mourad, Astrophys.\ Space Sci.\  {\bf 314}, 341 (2008);
    A.~M.~Abbassi, S.~Mirshekari and A.~H.~Abbassi,Phys.\ Rev.\ D {\bf 78}, 064053 (2008).

\bibitem{M2} 
    E.~C.~Vagenas, Mod.\ Phys.\ Lett.\ A {\bf 21}, 1947 (2006);  Int.\ J.\ Mod.\ Phys.\ D {\bf 14}, 573 (2005);         T.~Multamaki, A.~Putaja, I.~Vilja and E.~C.~Vagenas,Class.\ Quant.\ Grav.\  {\bf 25}, 075017 (2008).

\bibitem{M3} 
   V. C. de Andrade, L. C. T. Guillen and J. G. Pereira, Phys. Rev.Lett. {\bf 84}, 4533 (2000); S. L. Loi and T. Vargas,       Chin. J. Phys. \textbf{43}, 901 (2005); O. Aydogdu, Int. J. Mod. Phys. D \textbf{15}, 459 (2006); 
    O. Aydogdu and M. Salti,  Czech. J. Phys. \textbf{56}, 8 (2006); S. Aygun and I. Tarhan, Pramana: J. Phys.   \textbf{78}, 531 (2012);  P.~K.~Sahoo, K.~L.~Mahanta, D.~Goit, A.~K.~Sihna, U.~R.~Das, A.~Prasad and R.~Prasad, [arXiv:1409.6513 [gr-qc]].
\bibitem{Bardeen} J.~Bardeen. in Proceedings of GR5, Tiflis, U.S.S.R., 1968.
\bibitem{AG} E. Ay\'{o}n-Beato and A. Garc\'{\i}a, Phys. Rev. Lett., {\bf 80} 5056 (1998);  
              Gen. rel. grav., {\bf 31} 629 (1999).

\bibitem{Math} Mathematica 9.0.

\bibitem{Yangetl1} I-Ching Yang, Chi-Long  Lin and I. Radinschi, Int. J. Theor .Phys. {\bf 48} 248 (2009).
\bibitem{Yangetl2} I-Ching Yang, Chi-Long  Lin and I. Radinschi, Int. J. Theor .Phys. {\bf 48} 2454 (2009).


\bibitem{KSVLens} K.~S.~Virbhadra, D.~Narasimha and S.~M.~Chitre, Astron.\ Astrophys.\  {\bf 337}, 1 (1998);K.~S.~Virbhadra and G.~F.~R.~Ellis, Phys.\ Rev.\ D {\bf 62}, 084003 (2000); {\it ibid} {\bf 65}, 103004 (2002); C-M.~Claudel, K.~S.~Virbhadra and G.~F.~R.~Ellis, J. Math. Phys.{\bf 42}, 818 (2001); K.~S.~Virbhadra and C.~R.~Keeton, {\bf 79}, 083004 (2009); {\bf 77}, 124014 (2008); K.~S.~Virbhadra,Phys.\ Rev.\ D {\bf 79}, 083004 (2009).
                       
\end{references}
\end{document}